\documentclass[letterpaper,preprint,byrevtex,showpacs,prb]{revtex4-1}

\DeclareSymbolFont{lettersA}{U}{pxmia}{m}{it}
\DeclareMathAlphabet{\mathsfsl}{OT1}{cmss}{m}{sl}

\SetSymbolFont{lettersA}{bold}{U}{pxmia}{bx}{it}
\DeclareFontSubstitution{U}{pxmia}{m}{it}
\DeclareSymbolFontAlphabet{\mathfrak}{lettersA}
\DeclareMathSymbol{\piup}{\mathord}{lettersA}{"19}
\DeclareMathSymbol{\iTheta}{\mathalpha}{letters}{2}

\usepackage{amsmath}
\usepackage{amssymb}
\usepackage{multirow}
\usepackage{graphicx}
\usepackage{epstopdf}
\usepackage[hang,scriptsize]{subfigure}
\usepackage{bbm}
\usepackage{mathrsfs}
\usepackage{xcolor}
\usepackage[colorlinks=true,pdfstartview=FitV,linkcolor=blue,citecolor=blue,urlcolor=blue]{hyperref}

\makeatletter
\newcommand{\rmnum}[1]{\romannumeral #1}
\newcommand{\Rmnum}[1]{\expandafter\@slowromancap\romannumeral #1@}
\makeatother

\renewcommand{\vec}[1]{\boldsymbol{#1}}

\newcommand{\ii}{\mathrm{i}}

\newcommand{\blue}[1]{\textcolor{blue}{#1}}

\begin{document}

\title{Broadband Epsilon-Near-Zero Metamaterials with Step-Like
    Metal-Dielectric Multilayer Structures}

\author{Lei~Sun}
\author{Jie~Gao}
\author{Xiaodong~Yang}
\email[To whom all correspondence should be addressed. \\ Electronic address: ]{yangxia@mst.edu}
\address{Department of Mechanical and Aerospace Engineering, \\
    Missouri University of Science and Technology, \\
    Rolla, Missouri 65409, USA}

\begin{abstract}
The concept of the broadband epsilon-near-zero meta-atom consisting of layered stacks
with specified metallic filling ratio and thickness is proposed based on the Bergman
spectral representation of the effective permittivity.
The step-like metal-dielectric multilayer structures are designed to achieve
realistic broadband epsilon-near-zero meta-atoms in optical frequency range.
These meta-atoms can be integrated as building blocks for unconventional optical components
with exotic electromagnetic properties over a wide frequency range, such as the demonstrated
broadband directional emission and phase front shaping.
\end{abstract}


\maketitle

\section{Introduction}

Metamaterials are artificially structured materials with subwavelength components,
which can be designed to create extraordinary macroscopic electromagnetic properties
that do not exist in nature \cite{Veselago1968SPU,Pendry1999IEEE,Valentine2008Nat,
Alu2005PRE,Rainwater2012NJP,Pendry2006Sci,Schurig2006Sci,Liu2007Sci,Shin2009PRL,
Choi2011Nat,Yang2012NP}.
Among all kind of metamaterials, the metamaterials with near-zero permittivity
(epsilon-near-zero, ENZ) emerge into the focus of the extensive exploration due to
their anomalous electromagnetic features at microwave and optical frequencies
\cite{Enoch2002PRL,Silveirinha2006PRL,Silveirinha2007PRB,Liu2008PRL,Edwards2008PRL,
Alu2007PRB,Feng2012PRL,Cheng2012PRL,Silveirinha2007PRE,Alu2007OE,Alu2009PRL,Alu2010PRL}.
However, all ENZ metamaterials suffer from single frequency response for the near-zero
permittivity, which gives a huge disadvantage for practical applications.

In this paper, we further develop the theoretical strategy \cite{Sun2012JOSAB}
for mathematically designing the broadband ENZ metamaterials based on the
Bergman spectral representation \cite{Bergman1979PRB,Bergman1992SSP}
of the effective permittivity, and propose the realistic broadband
ENZ meta-atoms by applying the step-like metal-dielectric multilayer structures.
In quasi-static conditions, the Bergman spectral representation expresses the effective
permittivity of a composite medium in terms of a series of residue-singularity couples.
Therefore, the broadband ENZ metamaterial can be flexibly designed
in a dimensionless spectral space,
while the geometric information of the metamaterial, including the dimensions
and the filling ratio of each component, can be obtained through a proper inverse
problem.
According to this strategy, the step-like metal-dielectric multilayer structures are
proposed to realize the broadband ENZ meta-atoms in optical frequency range.
Functional optical devices constructed by the realistic broadband ENZ meta-atoms
are also demonstrated, and exotic electromagnetic properties of such devices are explored
including the broadband directional emission and phase front shaping.

\section{Design of Broadband ENZ Meta-atom}

As depicted in Fig.~\blue{1(a)}, the proposed broadband ENZ meta-atom is a two dimensional
three-layer stack with the dimensions of $100\,\mathrm{nm}\times25\,\mathrm{nm}$.
Each layer of the meta-atom is a metal-dielectric mixture with gold (Au)
as the metallic inclusion and silica ($\mathrm{SiO}_{2}$) as the dielectric host medium.
The permittivity of gold follows the simple Drude model
$\varepsilon_{\mathrm{Au}}=\varepsilon_{\infty}-\omega_{p}^{2}/(\omega(\omega+\ii\gamma))$
with dielectric constant $\varepsilon_{\infty}=5.7$,
plasma frequency $\omega_{p}=1.3666\times10^{16}\,\mathrm{rad}/\mathrm{s}$,
and damping constant $\gamma=3\times4.0715\times10^{13}\,\mathrm{rad}/\mathrm{s}$.
Here the damping constant is of the value three times higher than the bulk value in order to account
for surface scattering, grain boundary effects, and inhomogeneous broadening of the metal thin film.
The permittivity of silica is $\varepsilon_{\mathrm{SiO}_{2}}=2.1338$.
In quasi-static conditions, the Bergman spectral representation expresses the effective permittivity
of the meta-atom as
\begin{equation}
\label{eq:bergman-1}
    \varepsilon_{e}=\varepsilon_{\mathrm{SiO}_{2}}(1-F(s))
\end{equation}
with respect to the variable
$s=\varepsilon_{\mathrm{SiO}_{2}}/(\varepsilon_{\mathrm{SiO}_{2}}-\varepsilon_{\mathrm{Au}})$.
The spectral function $F(s)$ reads
\begin{equation}
\label{eq:bergman-2}
    F(s)=\sum_{i=1}^{3}\frac{F_{i}}{s-s_{i}},
\end{equation}
which is analytic everywhere with positive residue $F_{i}$ except for three simple singularities
$s_{i}$ confined as $0\leqslant s_{1}<s_{2}<s_{3}<1$.
Correspondingly, based on the geometric structure of the stack, including the filling ratio $f_{i}$
of the gold inclusion in each layer and the thickness $d_{i}$ of each layer, the effective
permittivity $\varepsilon_{x}$ of the whole stack reads
\begin{equation}
\label{eq:electric}
    \varepsilon_{x}(s)=\varepsilon_{\mathrm{SiO}_{2}}\left[\sum_{i=1}^{3}\frac{d_{i}/d}{1-f_{i}/s}\right]^{-1}
\end{equation}
with the total stack thickness $d=\sum_{i=1}^{3}d_{i}$.
Through the effective permittivity, the connection between the singularities and residues in the
spectral space [Eq.~\eqref{eq:bergman-1}] and the geometric structure of the stack in the physical
space [Eq.~\eqref{eq:electric}] can be set up as
\begin{equation}
\label{eq:inverse}
    \varepsilon_{\mathrm{SiO}_{2}}\left[1-\sum_{i=1}^{3}\frac{F_{i}}{s-s_{i}}\right]
    = \varepsilon_{\mathrm{SiO}_{2}}\left[\sum_{i=1}^{3}\frac{d_{i}/d}{1-f_{i}/s}\right]^{-1}.
\end{equation}
Clearly, the variable $s$ is only a function of the frequency $\omega$ as $s=s(\omega)$,
which implies the Bergman spectral representation on the left hand side of Eq.~\eqref{eq:inverse}
describes the effective permittivity of the stack only as a function of the frequency $\omega$.
Therefore, for designing the broadband ENZ meta-atom, it is simple to request that
\begin{equation}
\label{eq:reeps}
    \mathrm{Re}(\varepsilon_{x}(s(\omega_{j})))
    = \varepsilon_{\mathrm{SiO}_{2}}\mathrm{Re}
    \left[1-\sum_{i=1}^{3}\frac{F_{i}}{s(\omega_{j})-s_{i}}\right]
    = 0
\end{equation}
with respect to a series of specified frequencies $\omega_{j}$ $(j=1,2,3)$ in
a wide frequency range $[\omega_{a},\omega_{b}]$.
Regarding the mathematical structure of Eqs.~\eqref{eq:electric} and \eqref{eq:reeps},
the value of the singularity $s_{i}$ in Eq.~\eqref{eq:reeps} can be set as $s_{1}=0$
and $s_{i}=\mathrm{Re}(s(\omega_{i}))$ for $i=2,3$ with respect to another series of
specified frequencies $\omega_{i}$ within $[\omega_{a},\omega_{b}]$.
The specified frequency $\omega_{i}$ and $\omega_{j}$ can be simply arranged as an
equally spaced array based on the following confinement
\begin{equation}
    \omega_{a}=\omega_{j=1}<\omega_{i=2}<\omega_{j=2}<\omega_{i=3}<\omega_{j=3}=\omega_{b}.
\end{equation}
Therefore, the residue $F_{i}$ can be fully determined by Eq.~\eqref{eq:reeps}.
With the value of the residue $F_{i}$ and the singularity $s_{i}$,
the filling ratio $f_{i}$ and the thickness $d_{i}$ of the broadband ENZ meta-atom
can be achieved from Eq.~\eqref{eq:inverse}.
It is worth noting that this strategy is suitable for designing multilayered meta-atom for
realizing other constant permittivity over a wide frequency range in principle, and the
number of stack layers is not limited.
In order to confine the optical loss within a moderate range,
the operating frequency range of the broadband ENZ meta-atom in Fig.~\blue{1(a)}
is set between $\omega_{a}=537.42\,\mathrm{THz}$ and $\omega_{b}=589.62\,\mathrm{THz}$,
and the filling ratio $f_{i}$ of the gold inclusion and the thickness $d_{i}$ of each layer are
designed and listed in Table~\blue{1}.
\begin{table}[htb]
\label{tal:table-1}
  \centering
  \caption{Filling ratio of the gold inclusion and the thickness of each layer}
  \begin{tabular}{ccc} \\ \hline
    Layer &\quad $f_{i}$ &\quad $d_{i}\,(\mathrm{nm})$ \\ \hline
    1 &\quad 0.216329 &\quad 43.6884  \\
    2 &\quad 0.188523 &\quad 19.2691 \\
    3 &\quad 0.163955 &\quad 37.0425 \\ \hline
  \end{tabular}
\end{table}
According to the results, the realistic meta-atom,
shown in the right panel of Fig.~\blue{1(a)},
is also designed as a three-layer stack
with the same thickness $d_i$ of each layer,
but each layer is replaced by a gold-silica multilayer structure,
where the height $h_{i}$ of the gold is $h_{i}=f_{i}\cdot h$ $(i=1,2,3)$
with respect to the same filling ratio $f_{i}$ of the gold inclusion.
It is worth mentioning that in practical fabrication, the feature size of the fabricated thin
film has the accuracy of nanometer, and a detailed analysis on the robustness of the designed
parameters is presented in the Appendix.

The effective permittivity $\varepsilon_{x}$ of the effective medium meta-atom is plotted
in Fig.~\blue{1(b)} with the red curves standing for the real part and blue curves standing for the imaginary
part (the optical loss).
Moreover, the solid curves indicate the theoretical value of $\varepsilon_{x}$ based on Eq.~\eqref{eq:reeps},
while the dashed curves indicate the retrieved value, which is calculated from the scattering parameters
($S_{11}$ and $S_{21}$) of the effective medium meta-atom from the full-wave finite-difference time-domain
(FDTD) method.
Five equally spaced frequencies (ENZ frequency), at which the value of $\varepsilon_{x}$
equals to zero, are listed at the right bottom of the figure and marked as yellow dots on the curve.
ENZ frequencies~\rmnum{1},~\rmnum{3},~and~\rmnum{5}~represent the frequency $\omega_{j}$,
while \rmnum{2}~and~\rmnum{4}~stands for the frequency $\omega_{i}$.
It is clear that the FDTD retrieved results matches the theoretical results well,
with only small fluctuations caused by the inevitable phase variation of the reflected
electromagnetic wave from different stack layers in the effective medium meta-atom,
which is not included in the theoretical strategy.
Additionally, the value of the effective permittivity $\varepsilon_{y}$
of the effective medium meta-atom is $\varepsilon_{y}\approx2.8+0.001\ii$ in the
operating frequency range.

The upper panel of Fig.~\blue{2} gives the distributions of the absolute value of the local
electric field inside the effective medium meta-atom with a height of $250\,\mathrm{nm}$
at the five ENZ frequencies with the electromagnetic wave directly
illuminating from the bottom of the meta-atom,
while the lower panel of Fig.~\blue{2} shows the electric field inside the realistic meta-atom
array with the same height (including 10 realistic meta-atoms along the $y$-direction).
As indicated in Fig.~\blue{2}, the electric field is strongly confined and guided through
the meta-atom along different paths with respect to different ENZ frequencies,
due to the continuity of the electric displacement with near-zero permittivity.
The properly designed filling ratio of the gold inclusion in each stack layer ensures the zero
effective permittivity around each specified ENZ frequency.
Then the corresponding thickness of each stack layer modifies the interactions between the
adjacent layers, resulting in the broadband ENZ response across the desired frequency range.
Additionally, the similar distribution of the electric field inside the effective medium
meta-atom and the realistic meta-atom array, although disturbed by the inevitable scattering
caused by the realistic structures, informs that the realistic meta-atom also possesses a
similar broadband ENZ response that is coincident with the theoretical design.

\section{Analysis of Broadband ENZ Optical Devices}

Functional optical devices with broadband response corresponding to the broadband ENZ property
can be constructed by the realistic broadband ENZ meta-atom.
For example, Fig.~\blue{3(a)} schematically gives a prism made of the meta-atoms in air,
with an oblique upper surface of angle $\theta$ for realizing the directional emission
of electromagnetic waves.
The incoming electromagnetic wave with vertical wave vector
$\vec{k}_{\mathrm{in}}$ and Poynting vector $\vec{S}_{\mathrm{in}}$
can be collimated inside the prism as it crosses the lower interface from air,
while the refracted out-going electromagnetic wave with the wave vector
$\vec{k}_\mathrm{out}$ and Poynting vector $\vec{S}_\mathrm{out}$
will be normal to the upper interface.
Physically, the directional emission is resulted in the conservation of the tangential
component of the wave vectors at the upper interface between the prism and air, which can be
explained by the flat elliptical iso-frequency contour (IFC) of the meta-atom
compared with the circular IFC of air with radius $k_{0}=2\pi/\lambda$
in Figs.~\blue{3(b)} and \blue{3(c)}.
Clearly, the propagating direction of the out-going electromagnetic wave
will be quite close to the normal direction ($k'_y$-direction) of the upper
interface of the prism without the consideration of material loss [Fig.~\blue{3(b)}],
but slightly bend away when the material loss is taken into account [Fig.~\blue{3(c)}].

Correspondingly, Fig.~\blue{4} gives the simulation results of the directional emission
for prisms made of effective medium meta-atoms [Fig.~\blue{4(a)}]
and realistic meta-atoms [Fig.~\blue{4(b)}]
with an oblique interface of $\theta=15^{\circ}$
at the designed ENZ frequency of $537.42\,\mathrm{THz}$.
The distribution of the electric field amplitude $E_{x}$ is displayed,
with the power flow indicated by hollowed arrows.
Strong reflection occurs at the bottom of the prism due to the mismatched impedance
(analysis about the impedance matching to free space is given in the Appendix),
while the near-zero phase variation of the propagating electromagnetic wave inside the prism
and the small angle of refraction of the out-going electromagnetic wave at the upper interface
of the prism are observed due to the near-zero permittivity.
Furthermore, the variation of the angle of refraction with respect to different
frequencies [Fig.~\blue{4(c)}] reveals the broadband ENZ response of the realistic
meta-atom made prism,
through the comparison among the theoretical results obtained from the Snell's law (blue curve),
the simulation results of the effective medium prism (blue dots),
and the simulation results of the realistic meta-atom made prism (black circles).
It is worth mentioning that the deviations of the simulation results of the realistic
meta-atom made prism is due to the inevitable material loss and scatterings caused
by the internal structures,
and the average angle of refraction of the realistic meta-atom made prism
is about $6.3^{\circ}$ (red line) in the designed operating frequency range,
representing an effective group index of $n_{g}=0.42$.

Because of the relatively small phase variation caused by the near-zero permittivity,
the realistic broadband ENZ meta-atom can also be used to build optical devices for shaping the phase front
of electromagnetic waves as the $S$-shape lens example shown in Fig.~\blue{5}.
The example is demonstrated at the ENZ frequency of $537.42\,\mathrm{THz}$
for $S$-shape lenses made of effective medium meta-atom [Fig.~\blue{5(a)}]
and realistic meta-atom [Fig.~\blue{5(b)}],
with the results indicated by the distribution of the electric field amplitude $E_{x}$.
Clearly, the uniform phase pattern of the electromagnetic wave inside the $S$-shape lens
due to the near-zero permittivity leads to a phase front of the out-going electromagnetic wave
conformal to the upper interface of the $S$-shape lens,
although not exactly matches surface geometry especially for the realistic meta-atom made $S$-shape
lens, because of the inevitable scattering caused by the structure.
Besides, it is worth noting that the electric field amplitude is not uniformly distributed
in the space above the lens, because the electromagnetic wave will be focused by the concave part
of the $S$-shape lens, but diffused by the convex part.
Meanwhile, when the out-going electromagnetic wave just crosses the upper interface of the $S$-shape
lens, the simulation results of the average phase variation versus different frequencies [Fig.~\blue{5(c)}]
for $S$-shape lenses made of effective medium meta-atom (blue dots) and realistic meta-atom (black circle)
indicate a near constant phase variation in the designed operating frequency range,
implying the broadband ENZ response.
And the mean value of the average phase variation is about $0.7\,\mathrm{rad}$ (red line)
in the operating frequency range.

\section{Conclusions}

In conclusion, based on the Bergman spectral representation of the effective permittivity,
we have proposed a design to achieve the realistic broadband ENZ meta-atom consisting of
step-like metal-dielectric multilayered stacks in optical frequency range.
The broadband ENZ response is achieved due to the fact that the electric field can penetrate
through the meta-atom through different paths corresponding to different ENZ frequencies.
Functional optical devices with exotic electromagnetic properties,
including directional emission and phase front shaping,
over a wide frequency range of more than $50\,\mathrm{THz}$
have been constructed with the meta-atom,
and these unconventional devices can find many applications in optical
communications, imaging processing, energy redirecting, and adaptive optics.
Furthermore, our design can also be used in designing metamaterials with
other broadband constant parameters, such as magnetic permeability,
electrical conductivity, and thermal conductivity and diffusivity.

\section*{Acknowledgments}

This work was partially supported by the Department of Mechanical and Aerospace Engineering,
the Materials Research Center, the Intelligent Systems Center, and the Energy Research and
Development Center at Missouri S\&T, the University of Missouri Research Board,
and the Ralph E. Powe Junior Faculty Enhancement Award.

\section*{Appendix}

In order to achieve the best simulation results corresponding to the effective permittivity results
depicted in Fig.~\blue{1} of the main text, the values of the gold inclusion filling ratio and the layer
thickness for the realistic structures in Table~\blue{1} are calculated from the exact theoretical analysis
and they are up to six significant digits mathematically.
\begin{table}[htbp]
\renewcommand\thetable{A1}
\label{tal:table-A1}
    \centering
    \caption{Filling ratio of the gold inclusion and the thickness of each layer}
        \begin{tabular}{ccccc}
            \hline
            \multirow{2}{*}{Layer\quad} & \multicolumn{2}{c}{$f_{i}$} & \multicolumn{2}{c}{$d_{i}\,(\mathrm{nm})$} \\
            \cline{2-5}
            & Theoretical &\quad Approximated &\quad Theoretical &\quad Approximated\\
            \hline
            1  &\quad  0.216392    &\quad     0.22      &\quad    43.6884     &\quad  44   \\
            2  &\quad  0.188523    &\quad     0.19      &\quad    19.2691     &\quad  19   \\
            3  &\quad  0.163955    &\quad     0.16      &\quad    37.0425     &\quad  37   \\
            \hline
    \end{tabular}
\end{table}
\begin{figure}[htbp]
\renewcommand\thefigure{A1}
    \centering
    \includegraphics[width=7.5cm]{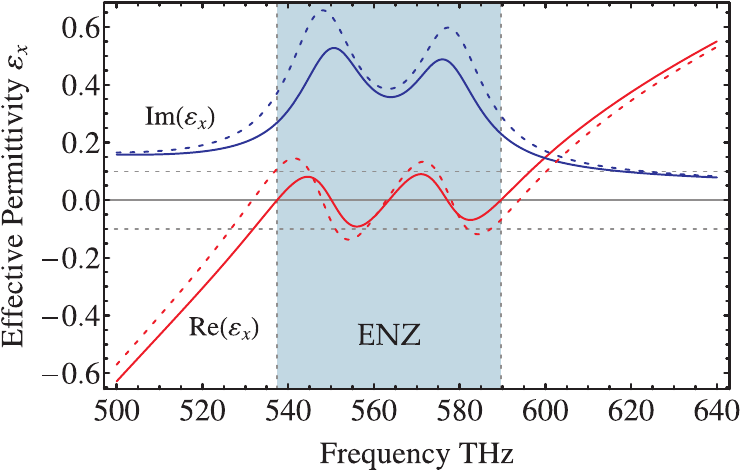}
    \caption{(Color online)
    Robustness analysis of the designed parameters in the realistic meta-atom
    with solid curves indicating theoretical results in the main text based on
    six-digit values of the fraction and the thickness,
    while the dashed curves indicating the approximated results based on two-digit values.}
    \label{fig:figA1}
\end{figure}

\noindent
In the practical fabrication, the feature size of the fabricated thin film has the accuracy of nanometer.
Regarding the robustness of the current design, a simple analysis is performed based on the theoretical design,
with the results indicated in the above Table~\blue{A1} and Fig.~\blue{A1}.
As shown in Table~\blue{A1}, the fraction and the thickness values are approximated into only two significant digits.
According to Fig.~\blue{A1}, the effective permittivity calculated with the approximated two-digit values
will only has a slight variation and small fluctuation, compared to the theoretical results obtained
from the exact six-digit values in the manuscript.
It is demonstrated that the performance of our design is quite robust to realistic fabrication tolerances.

On the other hand, the impedance marching with free space is of importance in practical applications.
According to the obtained effective permittivity shown in Fig.~\blue{1(b)}, the impedance
spectrum of the broadband ENZ metamaterial devices is plotted in Fig.~\blue{A2}.
\begin{figure}[htbp]
\renewcommand\thefigure{A2}
    \centering
    \includegraphics[width=7.5cm]{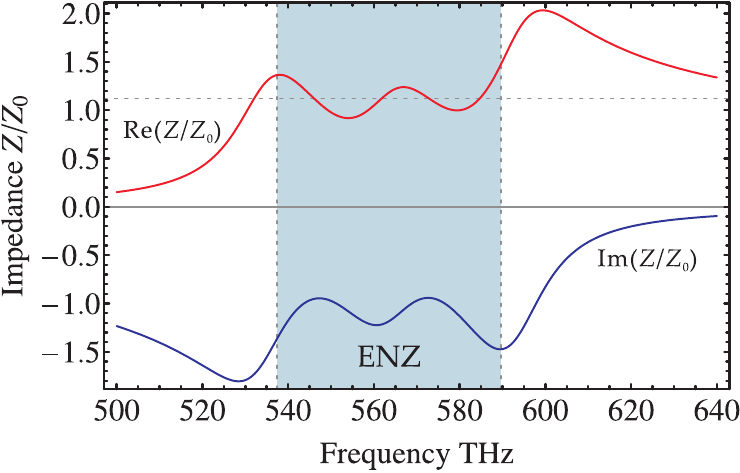}
    \caption{(Color online)
    The impedance of the designed broadband ENZ metamaterials}
    \label{fig:figA2}
\end{figure}
Within the broadband ENZ frequency region, the impedance also gives oscillatory
response with the definition of $Z/Z_{0}=\sqrt{\mu/\varepsilon_{x}}$, where $Z_{0}$ is the impedance
of free space.
Although the impedance is not perfectly matched with free space, the value is relatively low within
the ENZ frequency region, which is benefit from the existing of optical loss.


\newpage                 %
\section*{Figure Captions}%
\noindent
\textbf{FIG.~1}. (Color online)
The schematic diagram of the broadband ENZ meta-atom and the designed broadband
near-zero effective permittivity.
(a) The effective-medium meta-atom
with three homogenous layers composed by gold and silica,
and the realistic meta-atom with a step-like metal-dielectric multilayer structure.
(b)
The theoretical values (solid curves) and the FDTD retrieved values (dashed curves)
of the real part (red curves) and the imaginary part (blue curves)
of the effective permittivity $\varepsilon_{x}$ of the effective medium meta-atom,
with the ENZ frequencies (yellow dots) listed at the right bottom.

\vspace{5.0mm}
\noindent
\textbf{FIG.~2}. (Color online)
The distributions of the absolute value of the electric field in the effective medium
meta-atom (upper panel) and the realistic meta-atom array (lower panel) at different
ENZ frequencies,
corresponding to the incident electromagnetic wave indicated by the coordinates.

\vspace{5.0mm}
\noindent
\textbf{FIG.~3}. (Color online)
The schematic diagram and IFC analysis about the directional emission with
respect to the broadband ENZ response.
(a) The diagram of the directional emission through the ENZ prism
represented by the electromagnetic waves inside and outside the prism.
(b) The flat elliptical IFC (red curve for $\varepsilon_{x}=0.05$ and $\varepsilon_{y}=3.0$)
of the meta-atom under lossless condition.
(c) The elliptical-like IFC (red curve for $\varepsilon_{x}=0.05+0.5\ii$ and $\varepsilon_{y}=3.0+0.002\ii$)
of the meta-atom under lossy condition.
The IFC of air is plotted as blue circle.

\vspace{5.0mm}
\noindent
\textbf{FIG.~4}. (Color online)
Simulation results about the directional emission of the broadband ENZ prism with
the $15^{\circ}$-oblique upper interface,
represented by the distribution of the electric field amplitude $E_{x}$
and the power flow (hollowed arrows), for
(a) the effective medium made prism and
(b) the realistic meta-atom made prism.
(c) The angle of refraction (in degrees) as a function of the frequency,
based on the theoretical effective permittivity (blue curve),
and the simulation results of the effective medium made prism (blue dots)
and the realistic meta-atom made prism (black circles).
The average angle of refraction is $6.3^{\circ}$ (red line)
in the ENZ frequency range.

\vspace{5.0mm}
\noindent
\textbf{FIG.~5}. (Color online)
Simulation results about the phase front shaping of the broadband ENZ $S$-shape lens
with the radius of curvature of $3\,\mu\mathrm{m}$,
represented by the distribution of the electric field amplitude $E_{x}$, for
(a) the effective medium made $S$-shape lens and
(b) the realistic meta-atom made $S$-shape lens.
(c) The average phase variation (in radians) as a function of the frequency,
with respect to the effective medium made $S$-shape lens (blue dots)
and the realistic meta-atom made $S$-shape lens (black circles).
The mean value of the average phase variation is $0.7\,\mathrm{rad}$ (red line)
in the ENZ frequency range.


\clearpage
\newpage
\begin{figure}[htbp]
\renewcommand\thefigure{1}
    \centering
    \includegraphics[width=7.5cm]{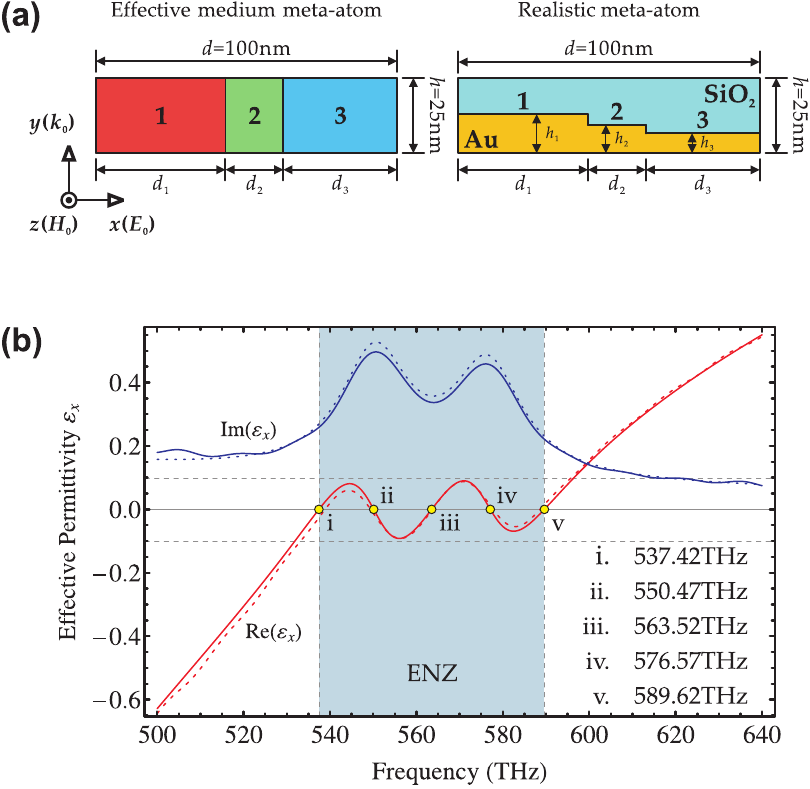}
    \caption{}
    \label{fig:fig1}
\end{figure}

\newpage
\begin{figure}[htbp]
\renewcommand\thefigure{2}
    \centering
    \includegraphics[width=7.5cm]{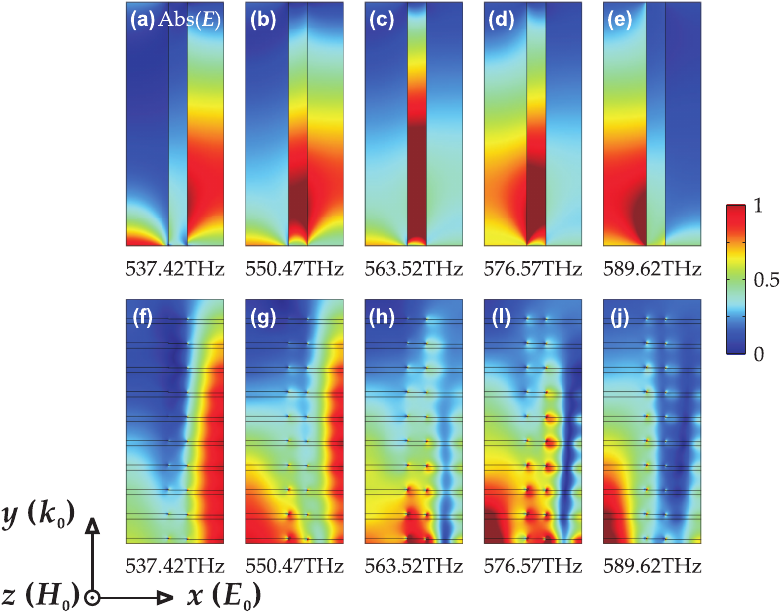}
    \caption{}
    \label{fig:fig2}
\end{figure}

\newpage
\begin{figure}[htbp]
\renewcommand\thefigure{3}
    \centering
    \includegraphics[width=7.5cm]{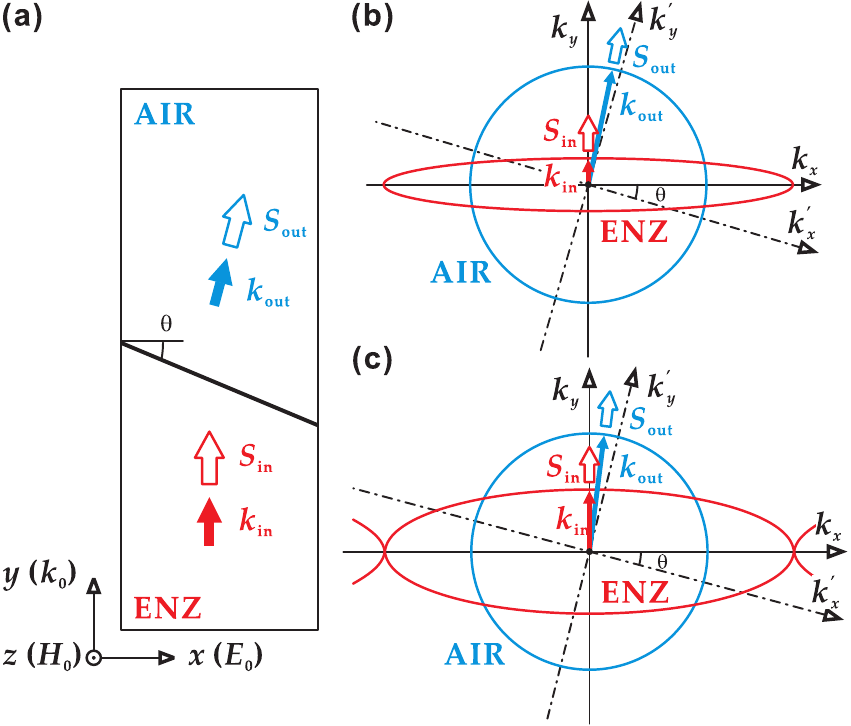}
    \caption{}
    \label{fig:fig3}
\end{figure}

\newpage
\begin{figure}[htbp]
\renewcommand\thefigure{4}
    \centering
    \includegraphics[width=7.5cm]{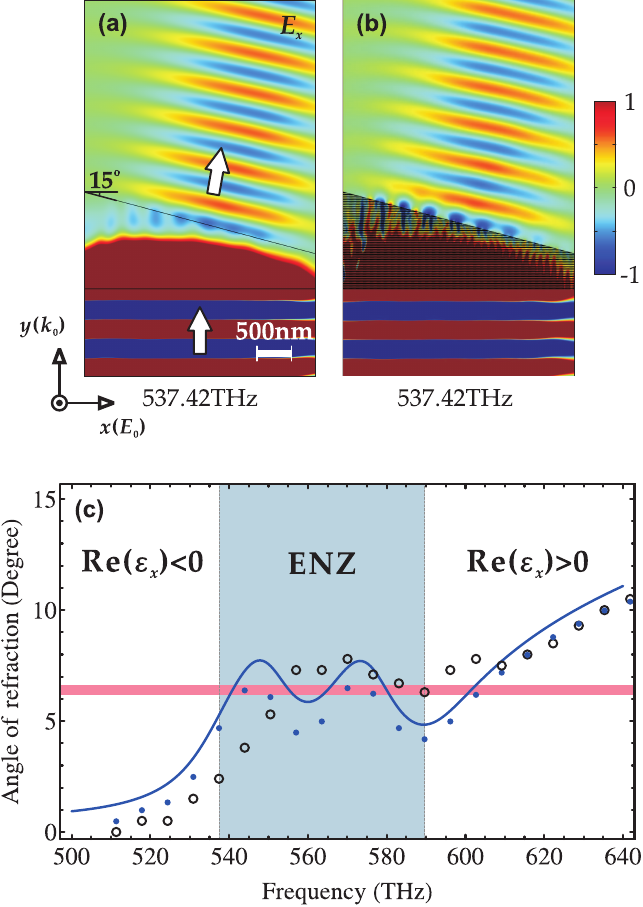}
    \caption{}
    \label{fig:fig4}
\end{figure}

\newpage
\begin{figure}[htbp]
\renewcommand\thefigure{5}
    \centering
    \includegraphics[width=7.5cm]{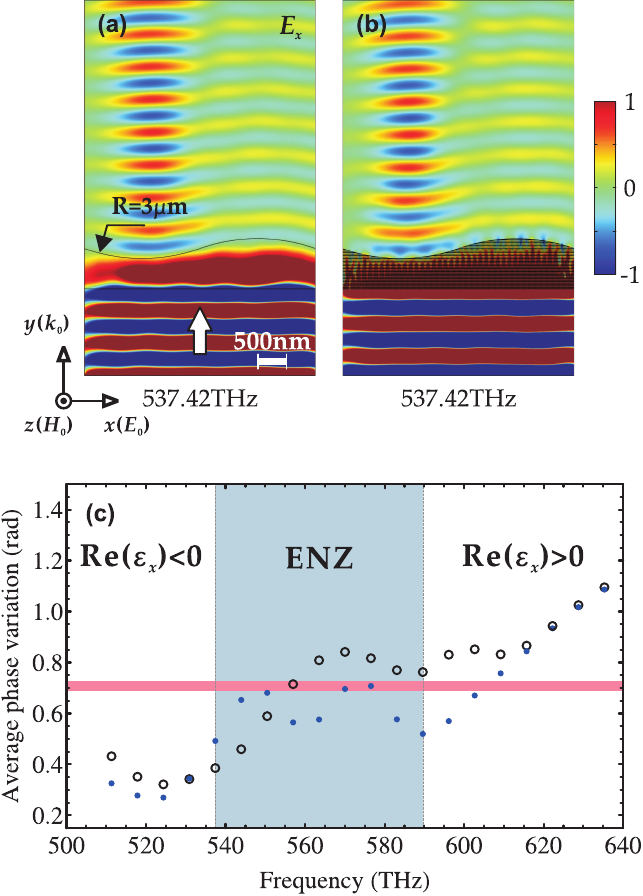}
    \caption{}
    \label{fig:fig5}
\end{figure}

\end{document}